# Bolder is Better: Raising User Awareness through Salient and Concise Privacy Notices

NICO EBERT
Zurich University of Applied Sciences, Winterthur, Switzerland

KURT ALEXANDER ACKERMANN
Zurich University of Applied Sciences, Winterthur, Switzerland

BJÖRN SCHEPPLER
Zurich University of Applied Sciences, Winterthur, Switzerland

This paper addresses the question whether the recently proposed approach of concise privacy notices in apps and on websites is effective in raising user awareness. To assess the effectiveness in a realistic setting, we included concise notices in a fictitious but realistic fitness tracking app and asked participants recruited from an online panel to provide their feedback on the usability of the app as a cover story. Importantly, after giving feedback, users were also asked to recall the data practices described in the notices. The experimental setup included the variation of different levels of saliency and riskiness of the privacy notices. Based on a total sample of 2,274 participants, our findings indicate that concise privacy notices are indeed a promising approach to raise user awareness for privacy information when displayed in a salient way, especially in case the notices describe risky data practices. Our results may be helpful for regulators, user advocates and transparency-oriented companies in creating or enforcing better privacy transparency towards average users that do not read traditional privacy policies.

**CCS Concepts** • Social and professional topics • Privacy policies • Security and privacy • Usability in security and privacy

**Additional Keywords and Phrases:** Privacy notices, privacy communication, privacy awareness, information saliency, information systems, experimental design

**ACM Reference Format:**

First Author's Name, Initials, and Last Name, Second Author's Name, Initials, and Last Name, and Third Author's Name, Initials, and Last Name. 2018. The Title of the Paper: ACM Conference Proceedings Manuscript Submission Template: This is the subtitle of the paper, this document both explains and embodies the submission format for authors using Word. In Woodstock '18: ACM Symposium on Neural Gaze Detection, June 03–05, 2018, Woodstock, NY. ACM, New York, NY, USA, 10 pages. NOTE: This block will be automatically generated when manuscripts are processed after acceptance.

## 1 INTRODUCTION

Researchers in the computer science and information systems community have established that privacy information, such as trust marks, privacy ratings, or privacy notices, can influence user behavior [20,22,23,41,46]. Less

*Please cite this preprint as*

*Ebert, N., Ackermann, K. A., & Scheppler, B. (2021, May) Bolder is Better: Raising User Awareness through Salient and Concise Privacy Notices. In Proceedings of the 2021 CHI Conference on Human Factors in Computing Systems*

straightforward, however, is the task of informing users about specific data practices. Privacy policies constitute the traditional approach of communicating data practices and have been studied extensively [40]. Since privacy policies primarily serve legal compliance requirements in allowing companies to limit their liability [5], privacy policies use formalized language that is frequently difficult to read and comprehend [9]. Moreover, many privacy policies are click-wrapped behind links. As a result, the corresponding policies are often ignored [5,7,21], and users may thus make wrong assumptions about the privacy practices [36] of the providers of websites and apps they use.

To remedy this, it was suggested that privacy policies should be supplemented with additional concise notices in natural language tailored to the relevant transaction context and target group [40]. Such "contextual privacy policies" [42] or "contextual privacy notices" [14:48] can be found on websites and on mobile apps of major companies. For example, Apple displays concise texts combined with a "data and privacy icon" on apps such as Apple Pay [49]. Another proposal to improve policy understanding is to increase relevance of notices by highlighting unexpected or risky practices [40], which is in the best interest of users, regulators or user advocates, but not necessarily of companies.

This paper addresses the question under what conditions concise privacy notices can best raise user awareness. To answer this research question, we conducted an exploratory online experiment in which we invited participants to test a new, fictitious fitness-tracking while varying the way privacy notices were displayed. Participants were asked to browse through the app and to indicate whether they would install it on their smartphones. Subsequently, they had to recall the privacy information displayed on the app. We chose a between-subjects factorial design with three levels of saliency of the notices (notices only available via click, notices shown exclusively on a dedicated screen, and notices embedded into existing screens) and two levels of privacy risk of the notices (privacy-friendly and privacy-intrusive).

In the following section, we focus on work related to privacy notices on conciseness, contextual relevance, saliency, and risk. We go on to explain the experimental methodology and report on our findings regarding information recall under the various conditions. Finally, we discuss the implications and suggest directions for further research.

## 2 RELATED WORK

### 2.1 Conciseness of Privacy Notices

When users are asked for reasons why they do not read privacy policies, they usually mention complex legal language and text length [34]. Indeed, a recent analysis of nearly 50,000 privacy policies on popular English-speaking websites regarding length and readability has confirmed that privacy policies are generally very long and hard to read [9]. In a longitudinal study of over 130,000 websites over a period of 20 years, it was recently found that privacy policies have doubled in size and the median reading level has risen [2]. Two experiments asking participants explicitly to read such policies have suggested that shorter privacy notices are preferable in informing users about data practices [15] and lead to higher information recall compared to traditional privacy policies [24].

While in the abovementioned experiments participants were explicitly asked to read the policies, other experiments did not include such an explicit reading assignment. In an experiment with a fictitious social network [16], only 26% of participants read the policy, and the average time they spent reading the policy text was found to be only one minute and 14 seconds for a lengthy policy of 7,977 words [36]. Given the assumption that individuals with a college degree are able to read approximately 280 words per minute [45], one would expect more than 28 minutes of reading time. In another experiment, participants were given a policy with only 451 words, which was displayed by default [44]. On average, participants spent one minute to read this relatively brief policy, which suggests that shorter texts are more likely to be read in more detail. Finally, in a recent experiment involving a fictitious social networking site, a shorter policy (approx. 300 words) outperformed a longer policy (approx. 2,000 words) in terms of information recall [31].

These findings are consistent with other research suggesting that shorter privacy statements are perceived as more comprehensible than lengthy, legalistic ones [39]. However, whether these notices are effective in a real-world situation, where users are not explicitly told to read them, has not yet been investigated.

## 2.2 Contextual Relevance of Privacy Notices

Nissenbaum developed the contextual integrity approach as a privacy theory. According to her theory, privacy is provided by an appropriate flow of personal information between players which is based on context-individual social norms [35]. Different contexts (e.g., healthcare vs. finance) can have different norms that influence which flows are considered appropriate and which are not. Nissenbaum considered the established procedural notice-and-consent mechanisms in online privacy based on privacy policies as "divorced from the particularities of relevant online activity" [35]. Accordingly, Schaub et al. [40] demanded that also privacy notices themselves should be contextualized and, for example, be embedded more effectively into contexts by focusing on contextually relevant information (e.g., specific unexpected risks). The terms "contextual privacy policies" [42] or "contextual privacy notices" [14:48] were coined. To increase the contextual relevance of policies, it has been suggested that they should focus on unexpected data practices [11], such as data sharing with third parties. Indeed, in a survey with self-reports of 500 older adults (50+), contextual relevance has been shown to be one of the factors that have predictive capacity with respect to reading behavior [38].

To identify what privacy information is potentially relevant in a given context, a recent study compared two contexts [8]. 642 people in two groups were asked about their privacy concerns and privacy information preferences in a loyalty card and fitness tracking context. In both contexts, users were most concerned about unauthorized secondary use and improper access. Also, information on the processing purpose was considered more important than information on automated decisions or the contact information of the data protection officer.

## 2.3 Saliency of Privacy Notices

Research has suggested that non-salient privacy notices fail to raise privacy awareness. When privacy notices are "hidden" behind a link, the click-wrapped content often goes unnoticed. Cate [4] mentioned the case of the Yahoo website in 2002, where only an average of 0.3% of the users read the click-wrapped policy. This proportion increased to only 1% after a public privacy firestorm [17]. In an experiment with a fictitious search engine, not one of the 120 participants clicked on the privacy policy link [16]. In another experiment, of the users asked to join a fictitious social network only 26% clicked on the policies [16]. Similar results were observed when participants were asked to participate in a survey and only 20.3% decided to click to see the privacy information [44]. This result compares to the results of an e-commerce experiment, where only 25.9% clicked on a policy link [22].

Saliency is an essential prerequisite of awareness not only on websites but also on mobile apps. Privacy notices are still predominantly displayed in ways that require users to click on a link. This also applies to prominent providers such as Google Playstore or Apple Appstore [50,51]. Presenting privacy information directly on an app screen rather than indirectly via a link, however, may improve users' awareness of data practices [25]. According to this hypothesis, research has suggested that users recall privacy notices better when they are directly and saliently displayed on the app itself [3]. However, empirical evidence for this hypothesis is currently missing.

## 2.4 Risk in Privacy Notices

In the domain of information security, the communication of risks to the user is well established. For example, Gates et al. [12] showed that a risk score displayed on the app store with additional permission-request-information had a significant influence on how users selected apps. Privacy warning labels have been discussed, too. In a 2007 experiment with 220 students and a stimulus website, privacy risks were displayed as warning labels, which increased the subjects' risk perception [28]. Furthermore, it has been suggested that privacy policies should focus on risks to increase their relevance [40].

Risks that may raise users' privacy concerns are related to a) unauthorized secondary use of the data (e.g. sharing data with third parties), b) improper access to data or c) an unfair collection of too much data (e.g. [30,43]), for instance.

To date, however, it is unclear whether concise privacy notices themselves may be effective in making users aware of potential risky practices. Our research addresses this question by varying the degree to which the data practices described in the privacy notices are risky.

# 3 METHODOLOGY

## 3.1 Hypothesis

Based on previous literature, our hypothesis was that concise privacy notices are recalled better when they are made salient (i.e., displayed exclusively on a screen or embedded into existing screens) as compared to making them available only via a click. Based on the assumption that information is more likely to be noticed when directly displayed in the relevant context, we also expected a better recall when the notices are embedded into existing screens close to related app information rather than being displayed exclusively.

Finally, in terms of risks, we expected privacy-intrusive practices to be recalled better than privacy-friendly practices because of potentially higher affective arousal in terms of indignation and negative surprise. We based this conjecture on previous findings in privacy and security research as well as on findings in cognitive science that show that unexpected and thus surprising information is recalled better [10].

## 3.2 Procedure

In order to test the effectiveness of concise privacy notices, we designed an online experiment[1]. Participants aged 16+ years were recruited from a German online panel ("meinungsplatz.de")[2]. Panel members received a general notification email ("You can participate in a new survey.") and were directed to the survey[3] upon clicking on the corresponding link. The cover story was that feedback was requested on an upcoming fitness tracking app ("Tractiv") for Android and iOS smartphones. A "fully functional demo version" of the app was displayed to all participants in the browser (see Figure 1) and participants could click through the app and move from one screen to the next. Participants could not participate in the study via mobile devices but only via laptops and desktops because a sufficient screen size was needed to be able to interact both with the prototype and the survey questions. The concise privacy notices were included in the app and had less than 60 words.

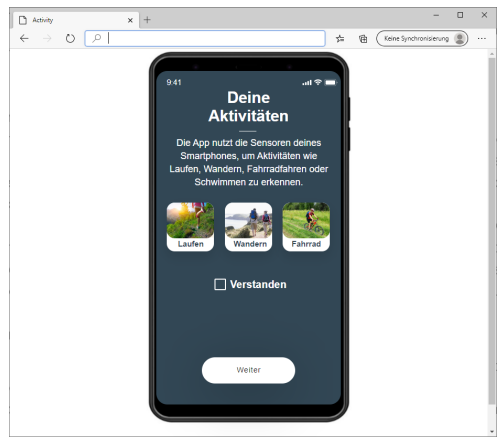

**Figure 1. The app prototype in the browser as seen by the users.**

The experiment comprised a 3 × 2 (saliency × privacy risk) between-subjects factorial design in which participants were randomly assigned to one of seven conditions (3 x 2 + 1 control). We varied three levels of saliency of the notices:

---

[1] The complete survey questionnaire, stimulus material, and experimental data are available at https://osf.io/kam3y/.
[2] In Germany, people over 16 are legally allowed to participate in online panels.
[3] Qualtrics was used as a survey tool. The fitness tracker app was embedded into the survey via iframe and built as a html/css/js website. A unique identifier per participant was provided by the panel and carried on through the survey including the app.

i) description of data practices only visible via clicking on a link ("click"), ii) description of data practices shown exclusively - i.e., as stand-alone information - by default on a dedicated app screen ("exclusive"), and iii) description of data practices embedded into several existing app screens as contextual information ("embedded"). Furthermore, we varied two levels of privacy risk of the notices: i) data practices with a low level of risk ("privacy-friendly"), and ii) risky practices ("privacy-intrusive"). When participants were done with browsing the app, the survey continued and participants were then asked to recall the data practices as a proxy for awareness. In the control condition, privacy notices were completely absent, such that subjects simply had to guess at the data practices employed, with their guesses indicating their expectations. Details regarding the measurement items are described in Appendix A.2.

The experiment was structured as follows:

- **Introduction and demographics**: Participants entered the survey and demographic information was collected.
- **Briefing and app presentation**: A briefing screen was displayed and afterwards the app was made accessible.
- **Distractor task, quality control questions and manipulation checks**: Additional questions were presented (e.g., sports activities, type of mobile devices) to ensure the recall questions did not only test sensory but also working memory [6,13]. This block also contained quality control questions and manipulation check questions.
- **Information recall**: Participants had to answer eight questions regarding the data practices of the app provider described in Table 1. The recall questions are described in Appendix A.2.
- **Debriefing and feedback**: On a debriefing screen, the actual purpose of the experiment and its authorship were revealed. Also, participants were given the opportunity to provide qualitative feedback.

According to the rules of the panel, participants were compensated with 0.6 € for their participation in the survey, which took them approximately 6 minutes to complete. To mitigate the risk of speeding or not engaging in the experiment, we measured the completion time, and we added a quality control question ("Click 'strongly disagree' if you have read this question.") in the distractor section. We used the completion time and the quality control question to exclude inattentive participants from the analyses.

The manipulation check questions served the purpose of assuring that the privacy-intrusive practices were, in fact, perceived as riskier than the privacy-friendly practices. That is, we asked participants to indicate their i) intention to use the app, ii) the degree to which they trusted the app, and iii) how risky they perceived the app to be.

Prior to the described data collection, we conducted a pre-test with 29 members of our faculty to assess the extent to which all questions were comprehensible and clear. Based on the results of the pre-test, some questions were modified to ensure comprehensibility.

## 3.3 Stimulus Material

In the fictitious fitness-tracking app, participants could click through a linear screen flow (for details see Appendix A.1). The screen flow started with a welcome screen (W) followed by four screens describing the main app features in a fixed order: automatic tracking of activities such as running, cycling, or hiking (A), tracking of participants' location to show activity on a map (L), voice control of the app (V) and music playback during activity (M). Finally, an in-app screen (I) with fictitious activity parameters was displayed (current activity type, duration of activity). In the following, we describe how the independent variables' factor levels were implemented in the app.

### 3.3.1 Saliency

We varied the three levels of saliency as follows:

- **Click**: In this treatment, a text with a link to a privacy notice was displayed (Figure 2, left). The position of the text and link within the screen flow (whether it was displayed at the beginning, in the middle, or at the end of the screen flow) was randomized to control for primacy and recency effects. However, it was placed at the bottom of the screen above the "Next" button in all instances. When a participant clicked on the "Next" button, he or she simply moved on with the screen flow. However, when a participant clicked the privacy link, a dedicated privacy notice screen was displayed (Figure 2, middle). This single screen described all data practices (Table 1) related to the abovementioned feature screens (A), (L), (V), and (M). By clicking "Next" on the notice screen, the participant moved on within the screen flow.
- **Exclusive**: In this treatment, the notice screen (Figure 2, middle) - which was only accessible via click in the first treatment - was displayed to all subjects by default and again contained all four privacy messages in a single screen. Again, the position of this screen within the screen flow (at the beginning, in the middle, or at the end) was randomized to control for primacy and recency effects. By clicking "Next" on the notice screen, the participants moved on within the screen flow.
- **Embedded**: In the final treatment, each one of the four privacy notices describing one data practice was displayed on a dedicated screen in conjunction with the related app feature (A), (L), (V), and (M) (Figure 2, right). For example, the message with privacy information related to activity tracking was displayed on screen (A) explaining the activity tracking feature. The message was always placed at the bottom of the screen above the "Next" button.

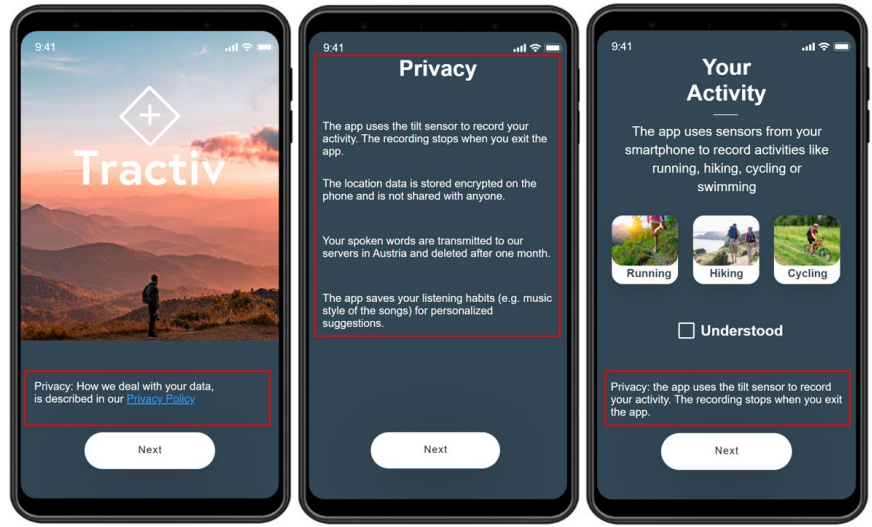

**Figure 2. The three different levels of saliency of privacy notices (translated from German): click (left), exclusive (middle), and embedded (right). The red boxes were not displayed to the users.**

### 3.3.2 Privacy Risk

The two levels of privacy risk (privacy-friendly vs. privacy-intrusive) are described in Table 1 for each of the four app features. Each data practice directly relates to a specific app feature. Across all treatments, the font and font-size of notice information were identical. The descriptions of both privacy risk levels had about the same length (54 and 55 words, respectively).

**Table 1: Data practices shown to participants (translated from German).**

| App feature | **Privacy-friendly** | **Privacy-intrusive** |
| --- | --- | --- |
| *Activity tracking (A)* | The app uses the tilt sensor to record your activity. The recording stops when you exit the app. | The app uses the microphone to record your activity. The recording remains active when you exit the app. |
| *Location tracking (L)* | The location data is encrypted and stored on the phone and is not shared with anyone. | The location data is stored in unencrypted form in the cloud and shared with our advertising partners. |

| App feature | **Privacy-friendly** | **Privacy-intrusive** |
|---|---|---|
| *Voice control (V)* | Your spoken words are transmitted to our servers in Austria and deleted after one month. | Your spoken words are sent to our servers in Russia and are not deleted. |
| *Music playback (M)* | The app saves your listening habits (e.g., music style of the songs) for personalized suggestions. | The app saves your listening habits (e.g., pirated songs) for distribution to companies. |

## 3.4 Subjects

The experiment was conducted in January 2020 over a period of nine days. 2,844 participants followed the invitation link. However, we had to exclude 442 respondents from analysis because the completion time was too short (< 160 seconds) or because of incorrect answers to the quality control question. From the 2,402 remaining participants, another 128 were excluded due to inexplicable browser activity or because the completion time was too long. Table 2 reports demographic information of the final sample of the 2,274 remaining participants across the seven conditions.

Table 2: Sample sizes and demographics across conditions (N = 2274).

|  |  | **Privacy-friendly** | | | **Privacy-intrusive** | | |
|---|---|---|---|---|---|---|---|
| *Conditions* | **Control** | **Click** | **Exclusive** | **Embedded** | **Click** | **Exclusive** | **Embedded** |
| *Participants* | 325 | 337 | 319 | 343 | 315 | 315 | 320 |
| *Gender (in percent)* | | | | | | | |
| Female | 45% | 47% | 50% | 51% | 47% | 49% | 48% |
| Male | 55% | 53% | 50% | 49% | 53% | 51% | 52% |
| *Age (years)* | | | | | | | |
| Mean | 47.85 | 46.91 | 47.27 | 47.06 | 48.42 | 47.26 | 47.00 |
| Sd | 14.35 | 14.84 | 14.56 | 14.46 | 14.35 | 14.61 | 14.89 |
| *Highest level of education (1 = none, 2 = primary, 3 = secondary, 4 = tertiary)* | | | | | | | |
| Median | 3 | 3 | 3 | 3 | 3 | 3 | 3 |

## 4 RESULTS

### 4.1 Recall Score

To evaluate participants' awareness of the data practices between the different conditions, participants had to answer eight questions on what data practices were employed by the app provider (two per data practice as described in Table 1). The recall score was then simply calculated as the number of correctly answered questions (0 - 8). Therefore, the maximum possible recall score was eight. Figure 3 (left) shows the mean recall scores across conditions. In the control condition, where no privacy information was displayed and participants had to guess, the scores refer to the number of guesses coinciding with the actual practices. Given that each of the eight recall questions offered five answer alternatives, one might expect an equal 20% chance of guessing the correct answer (i.e., an average of 1.6 correct questions). However, in the control condition, the guessing score for the privacy-friendly practices was higher (M = 2.80, SD = 1.29) and the score for privacy-intrusive practices (M = 1.35, SD = 1.07) lower than the expected value. This difference is statistically significant and corresponds to a medium-to-large effect size (t = 13.24, p < .001, n = 325, Cohen's d = .59). In other words, the participants' guesses in the control condition matched the privacy-friendly data practices significantly better than the privacy-intrusive data practices. That is, participants' expectations were significantly more consistent with privacy-friendly data practices than with privacy-intrusive data practices.

To be able to compare the recall scores across all conditions and to account for a different baseline probability of guessing correctly, we calculated an adjusted net recall score in the two treatment conditions concerning levels of privacy risk (Figure 3, right). The three treatments displaying privacy-friendly practices were adjusted according to the baseline for privacy-friendly practices, and the three treatments displaying privacy-intrusive practices were adjusted according to the baseline for privacy-intrusive practices.

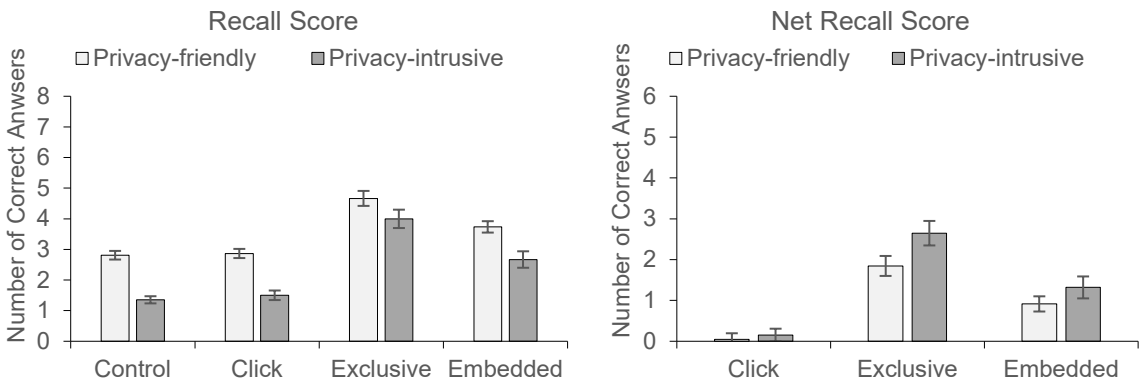

**Figure 3: Mean recall scores (left) and net recall scores (right) across conditions with 95% confidence intervals. Note that the number of correct answers in the control condition in the left figure does not indicate how well data practices were recalled (as privacy information was absent in the control condition) but rather how well the participants' guesses happened to match the data practices employed in the two conditions.**

### 4.1.1 Manipulation checks

Apart from the different recall baselines identified in the control condition ("guessing"), we used three additional items to verify that privacy-intrusive practices were, in fact, perceived as riskier than the privacy-friendly practices by means of manipulation check questions. That is, we asked participants to indicate their i) intention to use the app, ii) the degree to which they trusted the app, and iii) how risky they perceived the app to be. (Each item was measured with a 6-point Likert scale, 1=strongly disagree, 6=strongly agree; for details see Appendix A.2). We performed one-way ANOVAs with each item as a depended variable and the seven conditions as independent variables.

Additionally, Bonferroni-corrected post-hoc tests were performed to compare the conditions. We found a significant effect of treatments on the intention to use the app ($F_{6,2166}$ = 18.673, $p < .001$, n = 2173, Cohen's f = .226), a significant effect on perceived trust ($F_{6,1911}$ = 40.006, $p < .001$, n = 1918, f = .355), and a significant effect on perceived risk ($F_{6,1944}$ = 27.091, $p < .001$, n = 1951, f = .288). Post-hoc tests revealed that compared to the control group, only privacy-intrusive practices in the conditions embedded and exclusive had a significant impact (all $p = .000$). In the exclusive conditions, the mean score indicating the intention to use the app was reduced by 1.03, while in the embedded conditions it was reduced by .58 compared to the control group. The mean trust score was reduced by 1.33 in the exclusive conditions and .74 in the embedded conditions, and the risk score was increased by 1.13 in the exclusive conditions and .47 in the embedded conditions. In summary, our manipulation was successful, and the privacy-intrusive practices were indeed perceived as riskier than the privacy-friendly practices as indicated by a lower intention to use, less trust, and higher perceived risk, as can be seen in Figure 4.

Importantly, when only the click conditions are compared to the control condition, there are no statistically significant effects on intention to use (t = .71, p = .474, n = 932, Cohen's d = .04), trust (t =.68, p = .491, n = 788, d = 0.04) or risk (t = -.08, p < .932, n = 814, Cohen's d = .00).

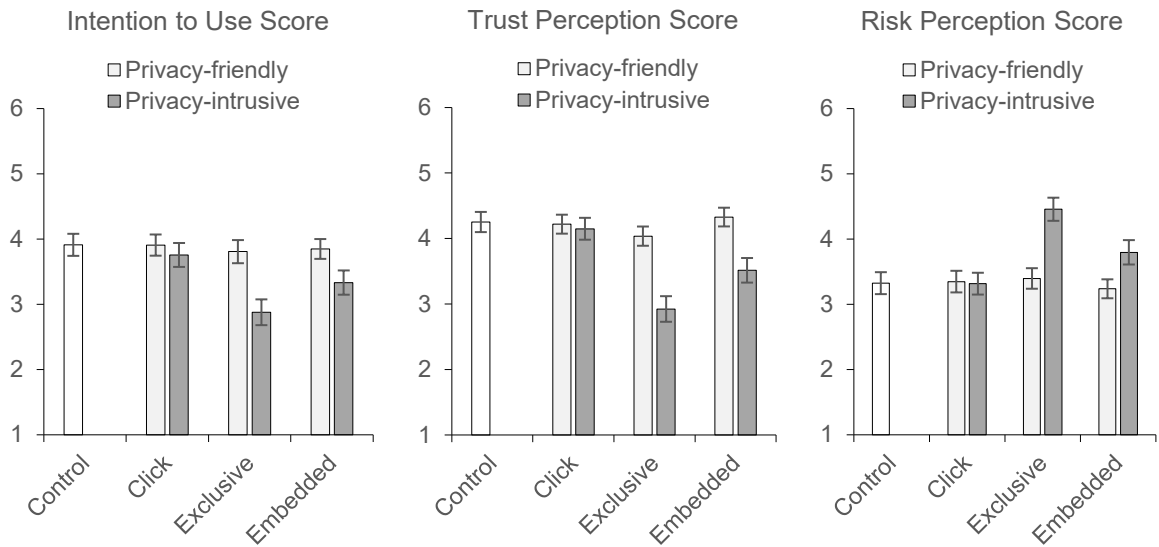

Figure 4: Intention to use, trust perception, and risk perception mean scores with 95% confidence intervals (1=strongly disagree, 6=strongly agree).

### 4.1.2 Regression model

To evaluate the effects that different factors have on data practice awareness, we performed a linear regression with the net recall score as dependent variable and the treatment variables saliency and privacy risk as independent variables while also including time spent with the fitness-tracker app ("app time", log-transformed), age, gender, and education level as control variables. Because the residual plot indicated heteroscedasticity, the heteroscedasticity-consistent standard errors were calculated. In particular, the common HC3 estimator was applied [18].

The parameter estimates of our model are reported in Table 3. The model is significant and accounts for roughly 25% of the variability in the individual net recall scores ($F_{9,1948}$ = 72.749, $p < .001$, n = 1948, $R^2$ = .252). While the variability of the score cannot be explained by age, gender, or level of education ($F_{5,1943}$ = .815, $p = .516$, $\Delta R^2$ = .002), $\Delta R^2$ was significant for the time the participants spent with the app ($F_{1,1942}$ = 285.367, $p < .001$, $\Delta R^2$ = .128) as well as for our treatment variables saliency and privacy risk ($F_{3,1939}$ = 105.713, $p < .001$, $\Delta R^2$ = .122). Therefore, our treatment variables account for about 50% of the total explained variance while the other roughly 50% of the total explained variance are accounted for by the amount of time that participants spent with the app - and so, potentially, also with reading privacy information. Importantly, as will be shown later, the amount of time people spent on the app - and hence also potentially on reading privacy information - is not independent of what treatment they were in.

To compare the net recall scores across conditions (Figure 3, right), we performed a two-way ANOVA (saliency x privacy risk) with Bonferroni-corrected post-hoc tests. The overall model was significant with a small effect ($F_{5,1943}$ = 76.787, $p < .001$, n = 1949, f = .197). There was a significant small main effect for saliency ($F_{5,1943}$ = 176.390, $p < .001$, f = .182), a significant but negligible main effect for privacy risk ($F_{1,1943}$ = 22.152, $p < .001$, f = .011), and a significant albeit negligible interaction effect ($F_{2,1943}$ = 4.675, $p = .009$, f = .005). Post-hoc tests showed that recall scores differ significantly (all $p < .001$) between the conditions click (M = .09, SD = 1.40), exclusive (M = 2.24, SD = 2.5) and embedded (M = 1.10, SD = 2.12). The extremely low average net recall score in the click conditions can be explained by the fact that only 16 out of 652 participants did, in fact, click on the policy link (2.5%).

Table 3: Linear Regression Results. *p < .001. a HC3 estimate

| Predictor | B | SEa | β | t | p | ΔR² |
|---|---|---|---|---|---|---|
| Constant | -5.058 | .884 | | -5.721 | .000 | |
| Age | -.005 | .003 | -.035 | -1.686 | .091 | |
| Gender | .097 | .088 | .022 | 1.099 | .272 | |
| *Education (null = None)* | | | | | | .002 |
| Primary | .268 | .805 | .053 | .332 | .682 | |
| Secondary | .267 | .804 | .059 | .331 | .682 | |
| Tertiary | .445 | .806 | .091 | .552 | .496 | |
| Ln(App Time) | 1.313 | .097 | .309 | 13.580 | .000 | .128* |
| *Saliency (reference level = Click)* | | | | | | |
| Exclusive | 1.886 | .109 | .395 | 17.321 | .000 | |
| Embedded | .912 | .097 | .193 | 9.450 | .000 | .122* |
| *Privacy risk (reference level = Privacy-friendly)* | | | | | | |
| Privacy-intrusive | .392 | .088 | .087 | 4.436 | .000 | |

#### 4.1.3 Comparison of net recall scores across conditions

The privacy risk manipulation also had differential effects on net recall scores for different levels of saliency. Due to the very low number of participants clicking on the policy link in the click conditions, there is no significant impact of privacy-intrusive vs. privacy-friendly data practices in these conditions (M = .46, SD = 1.40 vs. M = .15, SD = 1.41, p = 1). However, net recall scores differ significantly between privacy risk levels in the exclusive and embedded treatments (p < .001). Privacy-intrusive practices led to better recall compared to privacy-friendly practices in the exclusive conditions (M = 2.64, SD = 2.71 vs. M = 1.84, SD = 2.22) as well as in the embedded conditions (M = 1.31, SD = 2.45 vs. M = .91, SD = 1.75).

#### 4.1.4 Comparison of time spent and reading speed across conditions

Figure 5 shows the mean "extra time" in ln(seconds) that participants spent in the treatment conditions compared to the control condition, which contained no privacy information. Extra time was computed based on the app time of the conditions and can be interpreted as the additional time participants spent looking at privacy information. In accordance with our intuition, the results of the analysis indicate that in conditions with higher recall rates the participants also spent more time reading the privacy information.

We performed a two-way ANOVA (saliency x privacy risk) on extra time with Bonferroni-corrected post-hoc tests. The overall model is significant with a small effect ($F_{5,1943}$ = 9.240, p < .001, n = 1949, f = .153) that is accounted for by a significant small main effect of saliency ($F_{2,1943}$ = 21.836, p < .001, f = .149). The main effect of privacy risk was insignificant ($F_{1,1943}$ = 1.884, p = .170, f < 0.1) as was the interaction effect between saliency and privacy risk ($F_{2,1943}$ = .056, p = .813, f < 0.1). Regarding the main effect of saliency, post-hoc tests show significant differences (all p < .001) between the click conditions (M = -.01, SD = .51) and exclusive conditions (M = .17, SD = .53), on the one hand, and between exclusive and embedded conditions (M = 0.05, SD = .50), on the other hand. Within the conditions of saliency, there were no significant differences regarding privacy risk levels (all p > 0.5).

The average reading time for privacy-friendly (privacy-intrusive) policies was 1.7 seconds (3.2 seconds) in the embedded conditions and 7.8 seconds (11.2 seconds) in the exclusive conditions. Given the assumption that individuals are able to read approximately 250 - 280 words per minute or 4.1 - 4.6 words per second [45], one would expect a reading time of 11.5 - 12.9 seconds for the privacy-friendly policies and 11.7 - 13.2 seconds for the privacy-intrusive policies. Therefore, our data indicate that a substantial proportion of participants read most or even all of the privacy notices in the exclusive conditions, but only few appear to have done so in the embedded conditions and almost no one appears to have done so in the click conditions.

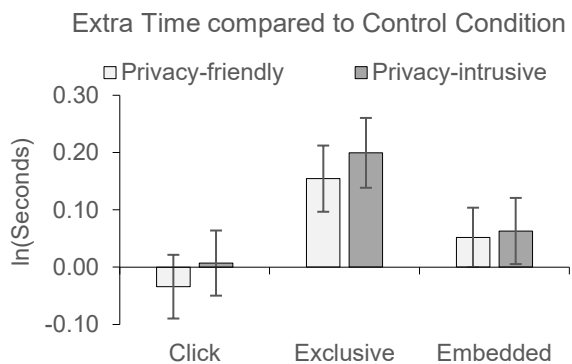

**Figure 5: Extra time spent on app in ln(seconds) compared to control condition with 95% confidence intervals.**

## 5 DISCUSSION

### 5.1 Findings

The goal of this exploratory online experiment was to investigate the degree to which concise privacy notices can raise user awareness. Furthermore, we wanted to explore the extent to which the saliency of privacy notices and the riskiness of data practices affect user awareness. Information recall was used as a proxy for users' privacy awareness. Our study deliberately focused on privacy notices with less than 60 words in simple language.

Our main finding is that concise privacy notices can raise user awareness of data practices also in real-world settings where participants are not explicitly told to read privacy notices. That is, our findings provide the first empirical evidence in support of the idea proposed by Schaub et al. [40] to provide concise notices complementary to traditional privacy policies. Our further results can be summarized as follows.

- The saliency level of the notices had a significant effect on awareness as measured by means of recall performance. Presenting privacy notices exclusively and prominently on a dedicated screen had a statistically significant and positive effect on recall compared to "hiding" notices behind a link that users have to click on, or embedding privacy notices on screens together with information on corresponding app features. To us, this finding was surprising as we assumed that information would be paid more attention to when it is embedded into the relevant context of a corresponding app feature - i.e., we assumed that a higher level of context integration would result in higher awareness. Our data indicate that the effect of exclusive presentation may be due to users spending more time on actually reading privacy information when it is explicitly, exclusively and - as a result - very saliently displayed to them as stand-alone information on a dedicated screen.
- Concerning the level of privacy risk, privacy-intrusive practices were less consistent with participants' expectations and – presumably because of that - better recalled than privacy-friendly practices, especially when they were presented exclusively on a dedicated screen. This finding is in line with our initial hypothesis as we conjectured that people do not expect providers to employ privacy-intrusive data practices and are, consequently, (negatively) surprised when realizing that a provider actually does so. The corresponding negative affective arousal that is supposed to be accompanied by such negative surprise may then lead to better recall and thus more awareness.

The first finding is in line with previous research on the perception of privacy information in general. When a privacy notice is "hidden" behind a link, the majority of people do not notice it (e.g., [36]). In contrast, when a privacy rating is made salient, it can have a significant effect on behavior (e.g., [1]). Our research indicates that, in terms of awareness, displaying privacy notices explicitly as stand-alone information on a dedicated screen seems to be superior to displaying it next to other information, even if this would ensure that the privacy information is more strongly contextually embedded and thus shown in conjunction with what it is actually relevant for. While Schaub et al. demanded relevant contextual privacy notices on the basis of the argument that contextually embedded notices are less disruptive [40], our findings suggest that disruption may in fact - at least to some extent - be necessary to make people notice privacy information in the first place.

Our second finding is in line with existing research on emotion and memory. Numerous studies have shown that emotionally arousing events are recalled better than affectively neutral events (e.g., [26,27,37]). In our case, these emotions are linked to unexpected and risky data practices. It is, therefore, indeed meaningful to highlight risky data practices in order to raise awareness when users do not expect such practices. Given that privacy notices are often "hidden" behind a link and thus hardly ever read by users in practice, false beliefs about generally benign data practices may perpetuate, and users may thus accept privacy policies and use apps on the premise of wrong assumptions.

Although it was not the primary aim of our study, our results also contribute to research investigating the relationship between privacy policy communication and trust in a corresponding service or institution. Lauer and Deng [29] as well as Wu et al. [48] found that consumers' trust in a company is closely linked to the perception of how well a company respects or does not respect customer privacy. However, based on the results of an experiment, Jensen et al. [22] suggested that the mere existence of a link to a privacy policy has a positive effect on users' confidence in a website. In contrast, Metzger [32] did not find a significant effect on different kinds of privacy assurances (e.g., policies, seals) on students' trust in an online retailer, even with a "weak" privacy policy that "gave notice that site visitors' data would be collected […] and said that this information could be passed to unauthorized third parties.". Our data indicate a significantly negative effect of privacy-intrusive data practices on intention to use, trust perception, and risk perception, given that these practices are presented in a salient way so that they are noticed. Therefore, out data support the linkage between the perception of privacy protection and trust. However, we do not find any significant differences regarding participants' intention to use the app nor the degree to which the app is trusted or perceived as risky between the control and the click conditions.

## 5.2 Limitations

Before considering implications for the design of privacy notices, some limitations of this study should be noted. First, the study was conducted in Germany, which generally has a high uncertainty avoidance culture compared to other countries [19]. Earlier research has shown that this characteristic is negatively correlated with privacy concerns [33], and thus the amount of attention paid to privacy information in our study may be underestimated. Furthermore, the app was used in a laboratory-like setting following a specific protocol (e.g., pre-defined screen flow) with limited app functionality and without requesting users to actually disclose personal data to use the app (i.e. we did investigate effects on user awareness and not on behavior). We also presented the app in the web browser and participants used a computer screen. We cannot rule out the possibility that a presentation of the app on a smartphone screen may have affected the results. Having selected a fitness-tracking app as stimulus may have had specific effects on recall (e.g. because specific privacy concerns were addressed). Furthermore, participants in our experiment had a monetary incentive to complete the survey. Nevertheless, we did not mention the research topic in the invitation e-mail, nor did we reveal further information on the purpose of the survey during the experiment. In fact, qualitative feedback given by subjects in the open text answers after debriefing revealed that some participants were surprised when they were informed about the actual purpose of the experiment. This indicates that subjects had not doubted the cover story when spending time with the app and had assumed that the app was real, which is in support of our findings' external validity. Nevertheless, a field study with a real app that runs un smartphones would be required to evaluate the degree to which our results hold in the field.

## 5.3 Implications and Further Research

Despite these limitations, our work has clear practical implications. We would encourage regulators, user advocates and privacy-friendly companies to make use of concise privacy notices to provide greater transparency and reach users who usually do not read traditional privacy policies. Moreover, to be most effective, concise privacy notices should be displayed exclusively as stand-alone information. Embedding privacy notices at the bottom of a screen combined with other contextually relevant information, which is similar to the way companies like Apple do it, appears to be less effective. As our research shows, users do not expect risky data practices and show high awareness when they encounter risky practices. Therefore, instead of purely enforcing traditional policies that do not mention risk or do not make them transparent, regulators may be able to enforce transparency over risks with these additional concise notices.

Concerning further work on privacy research, our contribution gives some potential directions. For instance, as already mentioned, our research should be replicated in a field study with a "real" app to measure the effect on user behavior in addition to user awareness. In addition, future studies may explore the extent to which other context factors (e.g., culture, social norms, particular type of app or website) and design properties (e.g., timing, length, position, color, language, content, additional icons) affect privacy awareness. For example, when to show the privacy notices during app use. Research on app permissions has shown that the effectiveness of permission requests is highly dependent on when the permission is requested during app use (e.g. ask-on-first-use of the app vs. asking during later usage with better contextualization of the request) [47]. Companies like Apple or Facebook have recently begun to use contextual privacy notices in combination with icons. However, it is yet unclear if this new approach is effective in terms of helping users to understand privacy information or if icons in general, or certain types of icons, distract users from relevant information and, in a worst case scenario, may even support the perpetuation of false assumptions about data practices.

## ACKNOWLEDGMENTS

This study was supported by the Hasler Foundation and the DIZH initiative of the Canton of Zurich. The authors would like to thank Mirjam Blumenstein for her contribution in building the app prototype and are indebted to Martin Degeling for pointing out the theory of contextual integrity to us.

# A  APPENDICES

## A.1  App Screenflow per Saliency Level

| Control | Click | Exclusive | Embedded |
| --- | --- | --- | --- |
| Welcome (W) | (W) w/ Link to notice screen* | (W) | (W) |
| Activity Tracking (A) | (A) | Default notice* | (A) w/ embedded notice |
| Location Tracking (L) | (L) w/ Link to notice screen* | (A) | (L) w/ embedded notice |
| Voice Control (V) | (V) | (L) | (V) w/ embedded notice |
| Music Playback (M) | (M) w/ Link to notice screen* | Default notice* | (M) w/ embedded notice |
| In-App Statistics (I) | (I) | (V) | (I) |
|  | * position randomized | (M) |  |
|  |  | Default notice* |  |
|  |  | (I) |  |
|  |  | * position randomized |  |

## A.2  Measurement Items (translated from German)

| Control variables | |
| --- | --- |
| Age | "How old are you?" (years) |
| Gender | "My gender is:" (1 = male, 2 = female) |
| Highest level of education | "My highest completed education is:" (1 = none, 2 = primary, 3 = secondary, 4 = tertiary, Classification according to International Standard Classification of Education) |
| App time | Duration of fitness-tracker app usage (in milliseconds) |
| Time | Duration of complete experiment (survey) (in milliseconds) |
| **Manipulation checks** | |
| Intention to use | "I could imagine using the Tractiv app shown here." (6-point Likert scale, anchored "strongly disagree" (1) and "strongly agree" (6) plus option "don't know") |
| Trust perception | "The Tractiv app is trustworthy." (6-point Likert, anchored "strongly disagree" (1) and "strongly agree" (6) plus option "don't know") |
| Risk perception | "It could be risky to use the Tractiv app." (6-point Likert scale, anchored "strongly disagree" (1) and "strongly agree" (6) plus option "don't know") |
| Recall questions (Questions and answers were randomized) | Which statement about recording your activity with sensors is true?<br>- The app uses the tilt sensor.<br>- The app uses the microphone.<br>- The app uses the accelerometer.<br>- The app uses the rotation sensor.<br>- The app uses the brightness sensor. |

Which statement about capturing your activity with sensors is true?
- Capture stops when you exit the app.
- Capture remains active when you exit the app.
- Capture stops automatically after 12 hours.
- Capture stops automatically after 24 hours.
- Capture stops automatically after 48 hours.

Which statement about your location data is true?
- The data is not shared with anyone.
- The data is shared with advertising partners.
- The data is shared with regulators.
- The data is shared with subsidiaries.
- The data is shared with other customers.

Which statement about your location data is true?
- The data is stored encrypted on the cell phone.
- The data is stored unencrypted in the cloud.
- The data is stored unencrypted on the cell phone.
- The data is stored encrypted in the cloud.
- The data is stored encrypted on the cell phone and encrypted in the cloud.

Which statement about the transmission of your voice inputs is true?
- The voice entries are transmitted to Austria.
- The voice entries are transmitted to Germany.
- The voice entries are transmitted to the European Union.
- The language input will be transmitted to China.
- The language input will be transmitted to Russia.

Which statement about the deletion of your language input is true?
- The language entries will be deleted after half a month.
- The language entries will be deleted after one month.
- The language data will be deleted after two months.
- The voice entries are deleted after twelve months.
- The voice entries will not be deleted.

Which statement regarding the storage of your listening habits is true?
- The listening habits are stored for advertising purposes.
- The listening habits are stored for personalized suggestions.
- The listening habits are stored for improvement purposes.
- Listening habits are stored for sharing with companies.
- The listening habits are stored for analysis purposes.

Which statement about storing your listening habits is true?
- The app stores e.g. illegally purchased songs.
- The app stores e.g. the song length of all songs.
- The app stores e.g. the artists of all songs.
- The app stores e.g. the titles of all songs.
- The app stores e.g. the music direction of all songs.